\documentclass[a4paper,11pt]{article}
\usepackage{pos}
\usepackage{gensymb}

\title{Search for enhanced TeV gamma-ray emission from Giant Molecular Clouds using H.E.S.S. }
 \ShortTitle{GMCs with H.E.S.S.}

\author*[a]{A. Sinha}
\author[b]{V. Baghmanyan}
\author[c]{G. Peron}
\author[a]{Y. Gallant}
\author[b,c]{S. Casanova}
\author[d]{M. Holler}
\author[c,e]{A. Mitchell}

\affiliation[a]{Laboratoire Univers et Particules de Montpellier, CNRS/IN2P3\\
  Universit\'{e} de Montpellier, F-34090 Montpellier, France}

\affiliation[b]{Institute of Nuclear Physics PAN, Radzikowskiego 152, 31-342 Kraków, Poland}

\affiliation[c]{Max-Planck-Institut f\"ur Kernphysik, P.O. Box 103980, D-69029 Heidelberg, Germany}

\affiliation[d]{Institute for Astro and Particle Physics,\\
University of Innsbruck, Innsbruck, Austria}

\affiliation[e]{Department of Physics,\\
ETH Zurich, CH-8093 Zurich, Switzerland}

\forColl{H.E.S.S.} % W/O "Collaboration"

\emailAdd{atreyee.sinha@umontpellier.fr}
\emailAdd{vardan.baghmanyan@ifj.edu.pl}
\emailAdd{giada.peron@mpi-hd.mpg.de}

\abstract{Cosmic Ray (CR) interactions with the dense gas inside Giant Molecular Clouds (GMCs) produce neutral pions, which in turn decay into gamma rays. Thus, the gamma ray emission from GMCs is a direct tracer of the cosmic ray density and the matter density inside the clouds. Detection of enhanced TeV emission from GMCs, i.e., an emission significantly larger than what is expected from the average Galactic cosmic rays illuminating the cloud, can imply a variation in the local cosmic ray density, due to, for example, the presence of a recent accelerator in proximity to the cloud.

Such gamma-ray observations can be crucial in probing the cosmic ray distribution across our Galaxy, but are complicated to perform with  present generation Imaging Atmospheric Cherenkov Telescopes (IACTs). These studies require differentiating between the strong cosmic-ray induced background, the large scale diffuse emission, and the emission from the clouds, which is difficult to the small field of view of present generation IACTs.

In this contribution, we use H.E.S.S. data collected over 16 years to search for TeV emission from GMCs in the inner molecular galacto-centric ring of our Galaxy. We implement a 3D FoV likelihood technique, and simultaneously model the hadronic background, the galactic diffuse emission and the emission expected from known VHE sources to probe for excess TeV gamma ray emission from GMCs.
}

\FullConference{37$^{\rm{th}}$ International Cosmic Ray Conference (ICRC 2021)\\
		July 12th -- 23rd, 2021\\
		Online -- Berlin, Germany}

\begin{document}
\maketitle

\section{Introduction}

The paradigm of origin and propagation of cosmic rays (CRs) is based on direct observations \cite{lipari2020shape} of CRs in the vicinity of the solar system \cite{Gabici2019}. However, in specific regions of the Galaxy, the level of CRs differs from the spectrum measured in the vicinity of Earth. Fermi-LAT \cite{Acero2016, yang}  observations show an up to $\sim$4 times higher density in the region around 4 kpc from the Galactic center, where the spectrum is also slightly harder. This challenges our current understanding on the propagation of CRs in the Galaxy. While a radial dependence of the propagation can give rise to such effects \cite{pothast2018progressive}, these effects can also be generated by the higher density of accelerators in the inner galactic regions \cite{giada}. 

Giant Molecular Clouds (GMCs) provide us with unique conditions for testing the density of CR in the Galactic disc. CRs interaction with the dense gas produces neutral pions, which in turn decay into gamma rays. Probing GMCs at different spatial points across our galaxy makes it possible to trace the distribution of galactic CRs point by point \cite{Casanova2010}. It is interesting to understand if the enhancement and hardening seen at different galacto-centric rings are localized in regions coincident with clouds or if they are characteristic of a larger region; it is only by directly probing GMCs that such information can be obtained. Such studies have been performed extensively with Fermi-LAT \cite{giada, yang, vardan} and enhanced cosmic ray density have been found near many clouds, especially in the inner galacto-centric rings.

Probing GMCs at Very High Energies (VHE) will be crucial to understand if the hardening seen at GeV energies continues in the TeV regime as well, constraining whether it is a local or global behaviour and if it is related to accelerators or to propagation effects. 

Studies of TeV emission from GMCs are, however, relatively rare. Enhanced emission has been reported from the galactic ridge by the H.E.S.S. collaboration \cite{ridge} spatially coincident with a complex of GMCs, while studies of high latitude clouds by HAWC have yielded upper limits consistent with expectations of a homogeneous \textquotedblleft sea\textquotedblright of CRs \cite{hawc}. Detection of such clouds with Imaging Atmospehric Cherenkov Telescopes (IACT) is challenging because it requires making a separation between the large residual hadronic background, the large scale diffuse emission, and the emission from the cloud, which is complicated due to the limited field of view of IACTs.

In this work, we use a full 3D likelihood method to separate the diffuse emission from the hadronic background, showing that this allows us to directly probe GMCs at TeV energies with a method similar to Fermi-LAT. As a proof of concept, we use this to investigate the region of cloud 877 from Rice et al.'s catalog \cite{rice}, located at 5.5 kpc from the Galactic Center. This cloud has been detected by Fermi-LAT to have a higher and harder spectrum than locally expected \cite{giada}. We show that this deviation continues at TeV energies, and thus, report on the detection of significant emission from the direction of this cloud at VHE.

\section{H.E.S.S observations}

The High Energy Stereoscopic System (H.E.S.S) is an array of five imaging atmospheric Cherenkov telescopes (IACTs) located in the Khomas Highland of Namibia, 1800 m above sea level. Results presented here contain data collected between 2004 - 2019, using the four 12m diameter telescopes, CT1-4, which have a field of view of $5 \degree$.
Observations used here have been carried out either as dedicated pointings on multiple individual objects or during the Galactic plane scan for the HGPS campaign. Calibration, reconstruction and $\gamma$-hadron separation were performed using the ImPACT chain \cite{impact} within the H.E.S.S. analysis package, and the counts and instrument response functions (IRFs) exported to a DL3 format as specified in the Gamma Astro Data Format\footnote{https://gamma-astro-data-formats.readthedocs.io/en/v0.2/} 

\section{3D Field of View likelihood technique}

A crucial element in any IACT analysis is the rejection of the residual hadronic background from $\gamma$-like cosmic ray induced showers. Standard background estimation techniques, e.g. the ring background or the reflected-region \cite{berge} rely on a measurement of the background in supposedly source-free regions within the observed field of view. However,  such approaches are poorly suited for our present case. They exclude faint emission at the level of the diffuse gas, and 
therefore are not suitable to trace molecular clouds, for which we expect a factor $\lesssim$ 10\% difference from the diffuse emission at TeV energies. A distinction between the of the hadronic component and the true diffuse emission is essential.

A three dimensional (3D) likelihood analysis provides us with a much more sensitive technique. While such techniques have been routinely used for high energy $\gamma$-ray data processing, its implementation in the field of IACT analysis is relatively recent \cite{luca, lars}. A spectro-morphological template model is constructed for the cosmic ray-induced background from archival runs with mostly empty fields of view. These runs are grouped according to zenith angle (and other observation conditions like optical efficiency, and an acceptance model is derived for each bin. Then, given an observation corresponding to certain parameters, a model background map is created which can then be directly fit as a separate component to the observed data. For details of background modelling, see \cite{lars}. 

The observed data is then described by a combination of many 3D (energy and 2 spatial dimensions) models, one for each expected emission component plus the constructed background model. The models are fitted to the data via a likelihood formalism and the significance of specific components determined by means of likelihood ratio tests. 

Such a likelihood analysis technique is implemented within the CTA Science tools, gammapy \cite{gammapy-icrc} . In this contribution, we use gammapy0.18\footnote{https://docs.gammapy.org/} \cite{gammapy-zenodo} to simultaneously model the residual hadronic background, the large scale diffuse emission and the emission from GMCs, and show that, for the first time, we achieve the detection of emission from the direction of a molecular cloud at TeV energies illuminated by background CRs.

\subsection{Analysis set-up}

The DL3 data are reduced following the standard gammapy analysis procedure. An energy range of $0.5 - 20$ TeV is chosen for the analysis, with events falling beyond an offset of 2.0\degree from the camera centre, or with energies below the peak of the background spectrum ignored.

\begin{figure}
    \centering
    \includegraphics[scale=0.3]{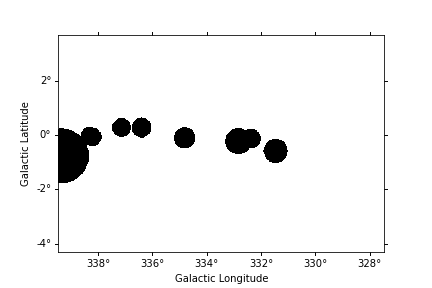}
   \includegraphics[scale=0.3]{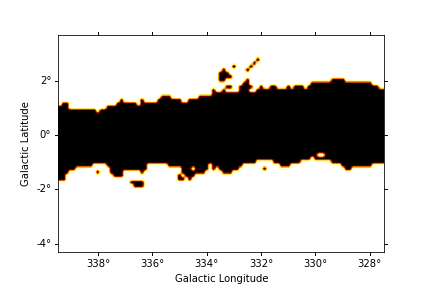}
   \includegraphics[scale=0.3]{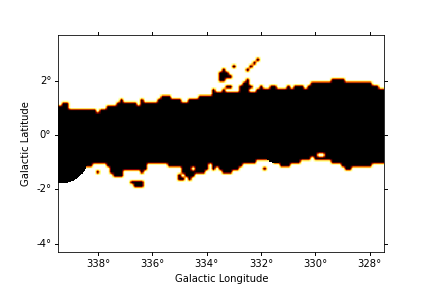}
    \caption{Exclusion mask used for the field of view normalisation: (left) A mask on the known HGPS sources, (middle) Mask constructed from dust column density for the diffuse emission, (right) The full mask, a product of the previous two.}
    \label{fig:exclusion}
\end{figure}

The background models, as constructed in the previous section, suffer from systematic fluctuations, and before stacking multiple runs, corrections to the model are required for each individual observation by fitting it outside exclusion region. For analysis of molecular clouds at low Galactic latitudes, it is crucial to make an optimal choice for the background exclusion region - exclude the maximal possible gamma-ray emission, while retaining enough off counts to do a do run by run background fitting. Hence, we choose an exclusion mask based on the Planck dust maps \cite{planck} which trace the interstellar gas. We excluded the dense gas, i.e., all pixels with a value of density above $2\times 10^{22}$ cm$^{-2}$.  Also, all known TeV sources are masked (Figure \ref{fig:exclusion}). A spectral correction to the background, (normalisation and tilt), is then fit for each run, and final stacked counts, background and IRFs constructed. 

\begin{figure}
    \centering
    \includegraphics[scale=0.32]{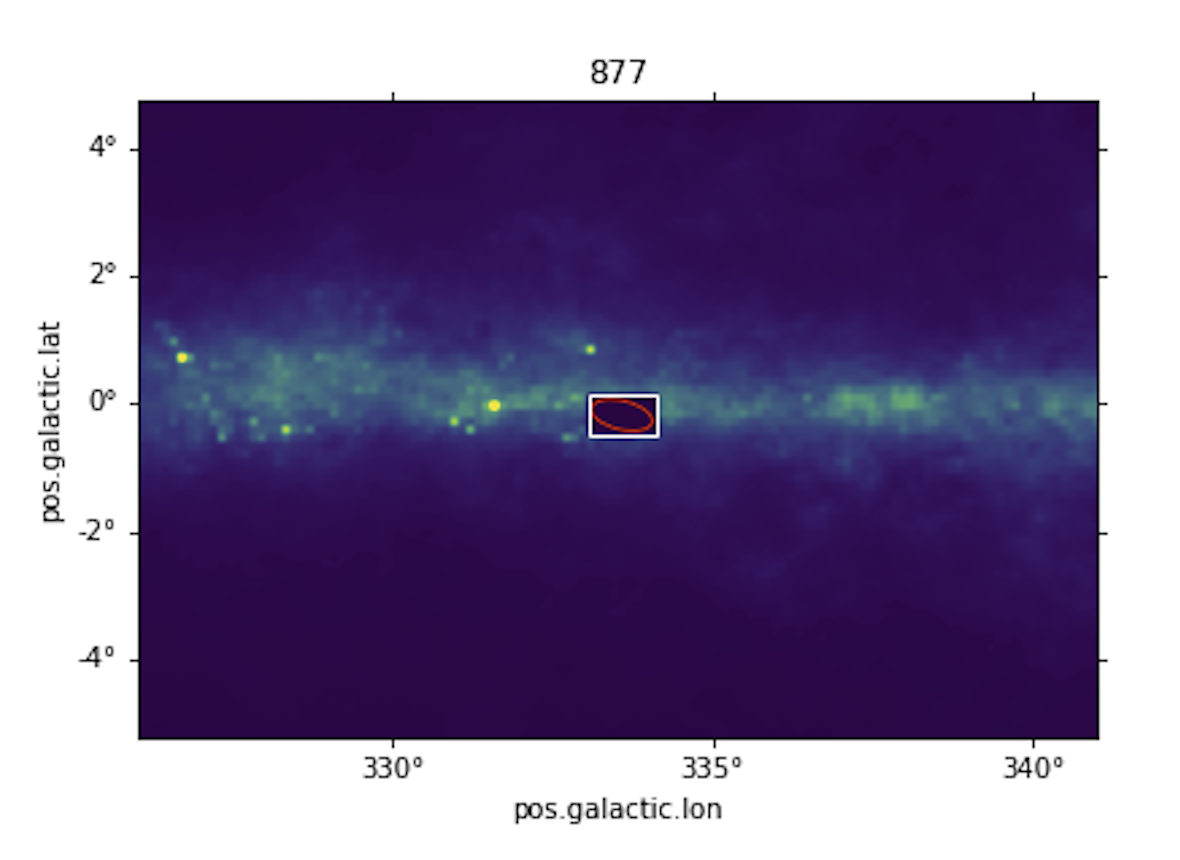}
    \includegraphics[scale=0.35]{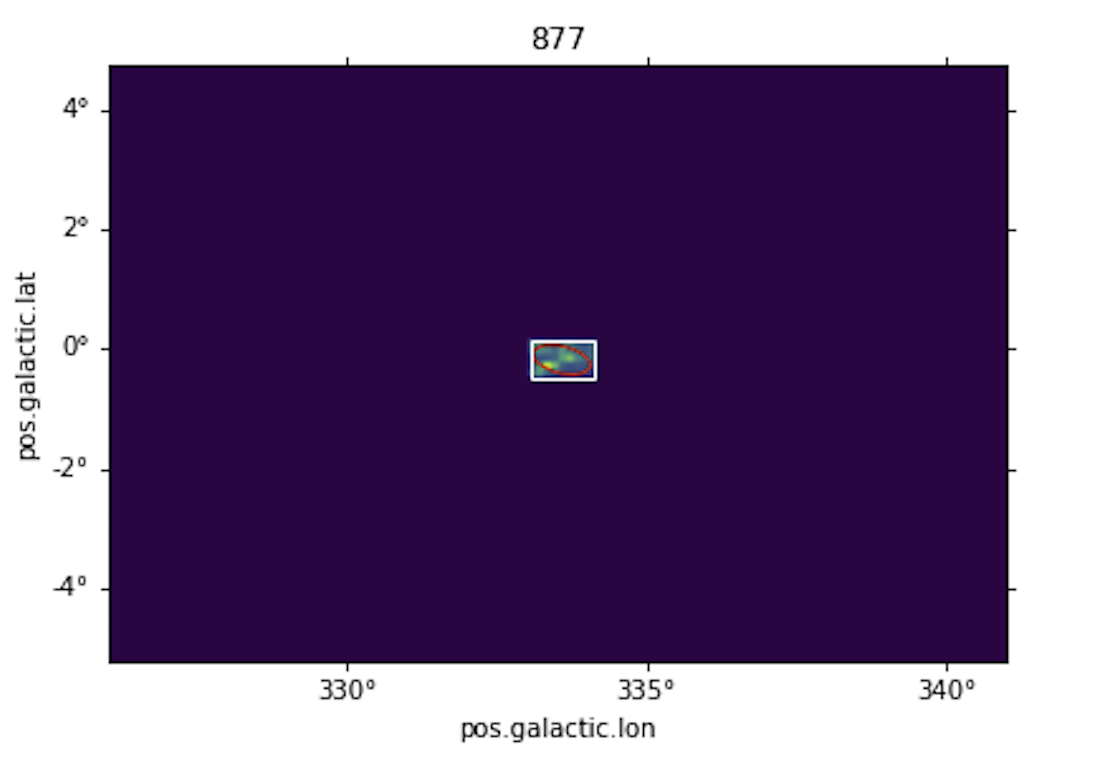}
    \caption{The templates used for modelling the diffuse emission and the cloud in the left and right panel, respectively. The elliptical size as quoted in \cite{rice} is plotted in red. }
    \label{fig:template}
\end{figure}

We use a technique similar to what is followed in \cite{yang,vardan,giada,hawc} to construct spatial templates for the pion decay emission. We use the dust 353 GHz opacity map which is a tracer of both molecular and atomic gas, and realize a rectangular cutout (Figure \ref{fig:template}) around the ellipse that represents the cloud extension, with $\sigma_{minor} = 0.22$ \degree, $\sigma_{major} = 0.48$ \degree, and position\_angle = 166 \degree as given in the clouds catalog \cite{rice}. 
To probe the emission from the cloud, we use the cut-out on the cloud as computed above.  The extracted spectrum refers to the dust template region which includes the cloud. Even though the extraction region formally includes the entire gas on the line of sight, in the case of cloud 877, the dominant contribution arises from a region in the velocity range coincident with the cloud which accounts for $\sim$60\% of the gas column. As demonstrated in \cite{peron2021probing}, observations of the entire column are a viable alternative in these cases, and can be successfully compared to the theoretical prediction that we have for the flux of galactic cosmic rays.

The rest of the dust map, with the cloud cut out, is used as a template for the large scale diffuse emission. Not modelling this correctly can lead to an over-prediction of the hadronic background, and in turn, an under-prediction of the flux from the direction of the cloud. Using a cut-out on the cloud allows us to analyze the diffuse emission as separate component and to extract the spectrum from there. The use of dust allows us to reduce the uncertainties related to the $X_{CO}$ conversion factor, the HI spin temperature, and the untraced (a.k.a. dark) gas. However, the downside is that a kinematic separation of the cloud is not possible, and the entire gas column in the given direction is taken as the cloud template. The maps are normalised such that they integrate to unity on the target geometry. The spectral model is assumed to be a simple power law, $F_{\gamma}(E) =  K(\frac{E}{E_0})^{-\alpha}$. We keep the known sources masked during the likelihood fit.

\section{Fermi-LAT analysis}
We performed an analysis of the region of the cloud 877 similar to the one in \cite{giada}, using the same spatial template applied in the H.E.S.S. analysis, namely a rectangular cutout based on dust templates. With respect to the published results \cite{giada}, based on 9 years of observations, we used here a collection of $\sim$ 12 years of data accumulated from August 4th 2008 to January 8th 2020. The standard quality cuts have been applied (\texttt{DATA\_QUAL==1 \&\& LAT\_CONFIG==1} and  $z_{max}=90^\circ$). We used \texttt{SOURCE} Pass8 data converted both at the front and at the back of the detector. Our model included all the sources of the 4FGL catalog. 
As done in \cite{giada}, we proceeded by fitting the diffuse sources first, with the parameters of the point sources fixed to the cataloged value. After a first fit, we proceeded with the optimization of the sources in the catalog. Then we performed a likelihood fit, freeing all the parameters of the sources within $3\degree$ from the center and leaving the normalization of the brightest sources (TS>1000) free. Finally, we investigated the residuals and included in the model any residual spots with TS >25. To investigate the influence of these new spots, we calculated the difference between the SED obtained for the cloud before and after adding the new sources. The difference was found to be lower than 2\% in each energy bin. 

%To be compatible with the H.E.S.S. spectral points, the Fermi-LAT analysis presented in \cite{giada} was redone using the dust templates described above (instead of CO+HI as in the published version, and a rectangular cutout for the cloud size instead of a square). Additionally, this uses 12 yrs of  accumulated data (vs. 9 yrs), and models all 4FGL sources in the Region of Interest (vs. 3FGL).

\section{Results}

\begin{figure}[h]
    \centering
    \includegraphics[scale=0.45]{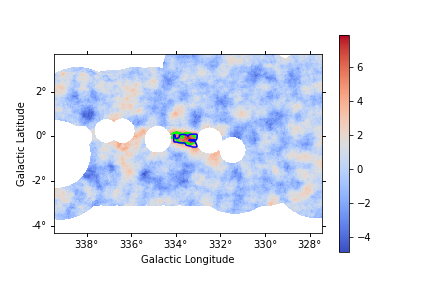}
    \includegraphics[scale=0.42]{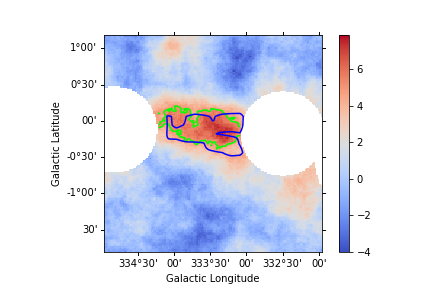}
    \includegraphics[scale=0.45]{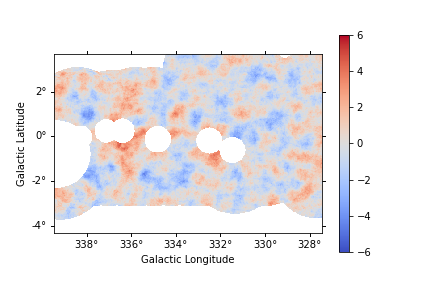}
    \includegraphics[scale=0.42]{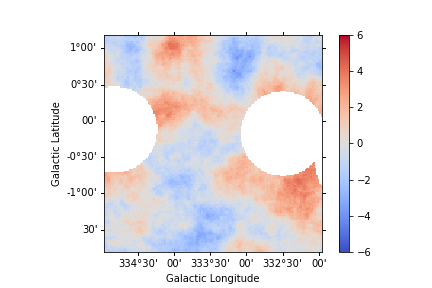}
    \caption{Top: Significance map of the cloud on a full RoI (left) and a cutout on the cloud region (right) with 4-sigma significance contour in green and 9e22 cm$ˆ{-2}$ dust column density contour in blue);
    Bottom: Significance maps after modelling the emission from the cloud}
    \label{fig:significance}
\end{figure}

Significant emission ($\Delta(TS)=101$) is detected from the direction of cloud 877.   In Figure \ref{fig:significance}, we show the significance map obtained before and after modelling the cloud, overlaid with $5\sigma$ significance contour lines and dust column densities. The H.E.S.S. significance contours well coincide with the gas contours, suggesting that the emission is truly correlated with the cloud.

\begin{figure}[h]
    \centering
    \includegraphics[scale=0.45]{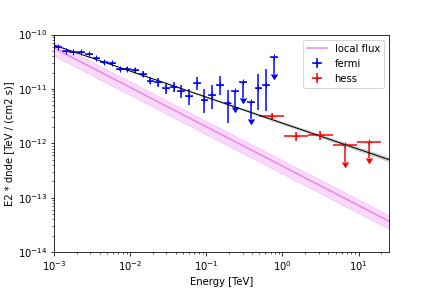}
    \includegraphics[scale=0.45]{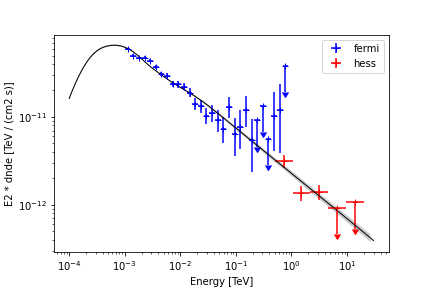}
    \caption{Joint fit between Fermi-LAT and H.E.S.S. spectral points using  {\bf Left}: An analytic power law spectral model. The gamma-ray spectrum expected from the local CR measurements is plotted in violet, the shaded region representing uncertainties in the cloud column density. {\bf Right} A pion decay model as implemented in \textsc{Naima}.}
    \label{fig:joint}
\end{figure}

The fitted spectrum ($\alpha = 2.73 \pm 0.16$) is well consistent with the extrapolation of the Fermi-LAT spectrum ($\alpha = 2.45 \pm 0.01$). In Figure \ref{fig:joint}, we show the results of a joint fit ($\alpha = 2.48 \pm 0.01$) between the H.E.S.S. and Fermi-LAT spectral points. With the opportune scaling, the expected flux of a cloud can be factorized in the emissivity $\phi(E)$ per H atom, and a factor $A\equiv \frac{M_5}{d_{kpc}^2}$, where $M_5$=M/10$^{5}$ M$_\odot$ and $d_{kpc}$ is the distance of the cloud in kpc, which accounts for the column density in the cloud. We followed \cite{giada} to compute the $\gamma$-ray emissivity, expected from a cloud illuminated by the local CR spectrum. We used a nuclear enhancement factor of 1.8 and a fit of the proton spectrum which interpolates the data of AMS-02 (up to $\sim$ 1 TeV) and DAMPE  (up to $\sim$ 100 TeV) \cite{lipari2020shape}.  
For cloud 877, $M_5=20$ and $d_{kpc} = 3.4 $, implying $A = 1.8$. Here the mass has been calculated for the cutout region used for the analysis, approximating the entire column to be located at the cloud distance, which is taken from \cite{rice}.

Interestingly, the TeV observation confirms the enhancement and hardening with respect to the local flux already reported for this region at GeV energies with Fermi-LAT.  The observed GeV-TeV emission is enhanced with respect to the local emissivity as derived from AMS-DAMPE spectral points, and significantly harder. With a joint fit of the Fermi-LAT and H.E.S.S. flux points using a pion decay 
model \cite{PionDecay} as implemented in the python package \textsc{naima} \cite{naima}, we % find the underlying proton population to have
estimate the differential cosmic-ray proton number density inside the cloud to be $\sim 1.5\times 10^{-17}$/GeV/cm$^3$ (at 1 TeV), with
an index of $2.58 \pm 0.01$.
% which is $\sim 5-6$ times the local density, and slightly harder.
This corresponds to an energy density of cosmic-ray protons with energies in the range 3--30\,TeV (roughly corresponding to the gamma-ray energy range of the H.E.S.S. flux points) of $1.0 \times 10^{-2}$\,eV/cm$^3$, which is $\sim 5-6$ times the local density.

\section{Conclusions}

We show that a full 3D likelihood analysis setup provides us with a powerful method of directly studying emission from GMCs. This technique is validated applied on Cloud 877 which is known to show excess emission in gamma rays, and we report on the first detection excess emission coincident with a passive GMC at VHE energies. This opens up exciting possibilities to probe the cosmic ray distribution across our Galaxy, and a detailed study with more clouds is in progress.

\section*{Acknowledgements}
\begin{tiny}
The support of the Namibian authorities and of the University of Namibia in facilitating the construction and operation of H.E.S.S. is gratefully acknowledged, as is the support by the German Ministry for Education and Research (BMBF), the Max Planck Society, the German Research Foundation (DFG), the Helmholtz Association, the Alexander von Humboldt Foundation, the French Ministry of Higher Education, Research and Innovation, the Centre National de la Recherche Scientifique (CNRS/IN2P3 and CNRS/INSU), the Commissariat à l'énergie atomique et aux énergies alternatives (CEA), the U.K. Science and Technology Facilities Council (STFC), the Knut and Alice Wallenberg Foundation, the National Science Centre, Poland grant no. 2016/22/M/ST9/00382, the South African Department of Science and Technology and National Research Foundation, the University of Namibia, the National Commission on Research, Science \& Technology of Namibia (NCRST), the Austrian Federal Ministry of Education, Science and Research and the Austrian Science Fund (FWF), the Australian Research Council (ARC), the Japan Society for the Promotion of Science and by the University of Amsterdam.

We appreciate the excellent work of the technical support staff in Berlin, Zeuthen, Heidelberg, Palaiseau, Paris, Saclay, Tûbingen and in Namibia in the construction and operation of the equipment. This work benefitted from services provided by the H.E.S.S. Virtual Organisation, supported by the national resource providers of the EGI Federation.
\end{tiny}

\bibliographystyle{JHEP}
\bibliography{references}

%% Full authors list (ONLY FOR COLLABORATIONS)
\clearpage
\section*{Full Authors List: \Coll\ Collaboration}
%
%\noindent \textbf{Note comment afterwards:} Collaborations have the possibility to provide an authors list in xml format which will be used while generating the DOI entries making the full authors list searchable in databases like Inspire HEP. For instructions please go to icrc2021.desy.de/proceedings or contact us under icrc2021proc@desy.de.\\
%
%\scriptsize
%\noindent
%first.author$^1$, 
%second.author$^2$, 
%third.author$^3$ % .... more names
%and 
%last.author$^{n}$ \\
%
%\noindent
%$^1$first.affiliation.
%$^2$second.affiliation. % .... more affiliation
%$^{m}$last.affiliation.

\scriptsize
\noindent
H.~Abdalla$^{1}$, 
F.~Aharonian$^{2,3,4}$, 
F.~Ait~Benkhali$^{3}$, 
E.O.~Ang\"uner$^{5}$, 
C.~Arcaro$^{6}$, 
C.~Armand$^{7}$, 
T.~Armstrong$^{8}$, 
H.~Ashkar$^{9}$, 
M.~Backes$^{1,6}$, 
V.~Baghmanyan$^{10}$
% ...

\medskip

\noindent
$^{1}$University of Namibia, Department of Physics, Private Bag 13301, Windhoek 10005, Namibia\\
$^{2}$Dublin Institute for Advanced Studies, 31 Fitzwilliam Place, Dublin 2, Ireland\\
$^{3}$Max-Planck-Institut f\"ur Kernphysik, P.O. Box 103980, D 69029 Heidelberg, Germany\\
$^{4}$High Energy Astrophysics Laboratory, RAU,  123 Hovsep Emin St  Yerevan 0051, Armenia\\
$^{5}$Aix Marseille Universit\'e, CNRS/IN2P3, CPPM, Marseille, France\\
$^{6}$Centre for Space Research, North-West University, Potchefstroom 2520, South Africa\\
$^{7}$Laboratoire d'Annecy de Physique des Particules, Univ. Grenoble Alpes, Univ. Savoie Mont Blanc, CNRS, LAPP, 74000 Annecy, France\\
$^{8}$University of Oxford, Department of Physics, Denys Wilkinson Building, Keble Road, Oxford OX1 3RH, UK\\
$^{9}$IRFU, CEA, Universit\'e Paris-Saclay, F-91191 Gif-sur-Yvette, France\\
$^{10}$Instytut Fizyki J\c{a}drowej PAN, ul. Radzikowskiego 152, 31-342 Krak{\'o}w, Poland\\
% ...

\end{document}